\documentclass[prb,twocolumn,superscriptaddress,showpacs]{revtex4}

\usepackage{graphicx}
\usepackage{bm}

\begin{document}

\title{Flat-Band Ferromagnetism in Organic Polymers Designed by a
Computer Simulation}

\author{Yuji Suwa}
\affiliation{Advanced Research Laboratory, Hitachi Ltd., Hatoyama,
Saitama 350-0395, Japan}
\author{Ryotaro Arita}
\affiliation{Department of Physics, University of Tokyo, Hongo,
Tokyo 113-0033, Japan}
\author{Kazuhiko Kuroki}
\affiliation{Department of Applied Physics and Chemistry,
University of Electro-Communications, Chofu, Tokyo 182-8585, Japan}
\author{Hideo Aoki}
\affiliation{Department of Physics, University of Tokyo, Hongo,
Tokyo 113-0033, Japan}

\date{\today}

\begin{abstract}
By coupling a first-principles, spin-density functional calculation with
an exact diagonalization study of the Hubbard model, we have searched
over various functional groups for the best case for the flat-band
ferromagnetism proposed by R. Arita et al.  [Phys. Rev. Lett. {\bf 88},
127202 (2002)] in organic polymers of five-membered rings.  The original
proposal (poly-aminotriazole) has turned out to be the best case among
the materials examined, where the reason why this is so is identified
here.  We have also found that the ferromagnetism, originally proposed
for the half-filled flat band, is stable even when the band filling is
varied away from the half-filling.  All these make the ferromagnetism
proposed here more experimentally inviting.
\end{abstract}


\maketitle

\section{Introduction}

Materials design to realize desired properties in condensed-matter
physics is becoming increasingly realistic.  One important factor is,
given high levels of computer performance, the computer simulation is an
essential part in designing.  Namely, computer simulations with first
principles calculations should be imperative in narrowing down, or even
pin-point, from a wide variety of candidate or designed materials for
the interesting properties and functions we search for.  Such an
approach is especially promising for molecules and nano-structures.
While the atom manipulation with scanning tunneling microscopy
(STM)\cite{Eigler,Shen,Hashi1,Taro2} provides new possibilities for
materials design, the design of molecules, especially polymers as we
focus on here, should be fundamental.

In this context organic molecules and polymers are of special interest,
since they have versatile structures and chemical properties that can be
wider than in inorganic materials.  Indeed, the discovery of a
conducting organic polymer by Shirakawa\cite{Shirakawa74,Shirakawa77}
kicked off intensive studies that are paving a new way to adopt organic
polymers or oligomers in realizing various functions as
molecular-electronics devices such as field-effect
transistors\cite{Garnier94} or light-emitting diodes\cite{Friend99}.

Organic ferromagnets have also attracted much attention as a challenging
target\cite{Allemand91,Rajca}.  In particular, organic magnets
consisting entirely of non-magnetic elements is of fundamental as well
as practical interests.  Ordinary ferromagnets consisting of magnetic
elements exploit electrons in d- or f-orbitals that are strongly
interacting.  Can magnetism arise in p-electron systems that are weakly
interacting?  One theoretical possibility is to apply the flat-band
ferromagnetism proposed by Mielke and by Tasaki\cite{Mielke1,Tasaki1}.
This arises as an effect of electron-electron repulsion when the
(one-electron) band structure contains a dispersionless band.  The
mechanism is interesting in many ways, but essential features are,
first, the system is totally distinct from the ``narrow-band limit'' in
textbooks, since the magnetism occurs when the transfer between
different sites is finite.

Second, this is a band ferromagnetism rather than a magnetism arising
from the exchange interaction between localized spins.  So, this
mechanism should be a good candidate for designing organic ferromagnets.
While there are also other possible mechanisms such as the
intra-molecular Hund's coupling for spins in
p-orbitals\cite{Allemand91}, the flat band ferromagnetism is
advantageous in that spins are carried by itinerant electrons which can
be utilized for spin injectors in spintronics.

Here a theoretical remark on the flat-band ferromagnetism is due.  The
flat-band ferromagnetism was first proposed by Lieb,\cite{Lieb} who
considers bipartite lattices (consisting of two sublattices) that have
different numbers of A sublattice sites $N_A$ and B sublattice sites
$N_B$.  Lieb proved that, when we switch on the electron-electron
repulsion (assumed to be short-ranged, so that we take the Hubbard
model), the ground state should then be ferr{\it i}magnetic with the
magnetization $\propto N_A-N_B$.  Quantum chemically, this model
contains $N_A-N_B$ non-bonding molecular orbitals, so is similar to
Mataga's model.  Shima and Aoki\cite{Shima93} then proposed a systematic
way to realize such systems as superhoneycomb structures.
Mielke\cite{Mielke2,Mielke3,Mielke4} and independently
Tasaki\cite{Tasaki4} then constructed other flat-band ferromagnetism,
which is distinct from Lieb's in that the flat band is constructed from
quantum mechanical interference between the nearest-neighbor and further
transfers (so the system is necessarily non-bipartite) and the ground
state is now ferromagnetic.

To be more precise, the lattice is required to satisfy what is called
the ``local connectivity condition''.  Namely, the flat band is by no
means a sufficient condition for ferromagnetism, and the magnetism
arises for special lattices on which adjacent ``Wannier'' orbitals {\it
have to} overlap with each other no matter how they are combined to
minimize each orbit size.  So, while ordinarily a flat band implies a
disjointed set of orbits, they are connected in Mielke-Tasaki lattices.
This is intuitively why spins align to lower the repulsion energy from
Pauli's principle.  Rigorous proof for the ferromagnetic ground state
has been given for the half-filled flat band.

In actual materials, it is very hard to realize perfectly flat bands,
and to make them perfectly half-filled.  For the former, a rigorous
proof (for Mielke-Tasaki lattice)\cite{Tasaki3} and numerical
simulations\cite{Sakamoto,Arita98} showed that a slight band dispersion
does not destroy the magnetism. For the latter, there are also a
proof\cite{Mielke1} and a numerical simulation\cite{Sakamoto} which
revealed that small deviations from half-filling are allowable for the
ferromagnetism.

In the context of this background, what kind of organic materials are
promising for realization of the flat-band ferromagnetism?  The hardest
part is to realize the connectivity condition.  In general we have to
consider complex lattices such as Kagom\'{e}, or lattices having
distant-neighbor transfers.  However, the condition is easier to satisfy
on one-dimensional polymers.  For example, Mielke-Tasaki model is
realized as a chain of triangles.\cite{Penc,Sakamoto,Arita3} So we can
concentrate on polymers.

Next comes the choice of the monomers that should be polymerized.
Even-membered rings such as benzene are disadvantageous in that
antiferromagnetic order tends to occur, which should dominate over the
ferromagnetism.  So we should opt for odd-membered rings, and the above
example of the chain of triangles is indeed an example of this.  Since
triangular molecules are scarce, we can conclude that we should focus on
polymers based on five-membered rings, such as polypyrrole,
polythiophen, etc.

For the computer design, the purpose of the present paper, there are two
tasks: (i) to search for the materials that realize the flat band {\it a
la} Mielke-Tasaki.  Since the chain of odd-membered rings is by no means
a sufficient (nor necessary) condition for the flat band, this is an
important part.

(ii) Next, we should do a first-principles (spin density functional)
band calculations to confirm (a) the flatness of the band and the
ferromagnetism in the ground state, and (b) whether the magnetism can
indeed be interpreted as the Mielke-Tasaki mechanism, for which we have
to evoke an electron-correlation calculation (exact diagonalization of
the Hubbard model here).  (c) We also address the question of how the
departure from half-filling affects the magnetism in the flat band.

For the first-principles calculation, we have employed the generalized
gradient approximation\cite{PBE96} (GGA).  Depending on the purpose of
each calculation, we have performed either the spin-unpolarized case
based on the density functional theory (GGA-DFT) or the spin-polarized
one based on the spin density functional theory (GGA-SDFT).  We used the
plane-wave-based ultrasoft pseudopotentials\cite{Vanderbilt,Laasonen}.
The energy cutoff was taken as 20.25 Rydberg.  The convergence criterion
of the geometry optimization was that all of the forces acting on each
atom were within $1\times10^{-3}$ Hartree/a.u.

The plan of this paper is as follows.  In section~\ref{SecExample}, we
describe the search for various materials and show some examples which
exhibit possibilities of ferromagnetism. In section~\ref{SecPAT}, we
show the result for the most promising material, poly-aminotriazole, and
discuss how the theory for the flat-band ferromagnetism works on this
particular material. In section~\ref{SecDiscuss}, we discuss how the
ferromagnetic state is robust against the deviation of the band filling
from the half-filling.

\section{Examined materials}
\label{SecExample}

\subsection{Flat bands in the polymer of five-membered rings}
\begin{figure}
\begin{center}
\includegraphics{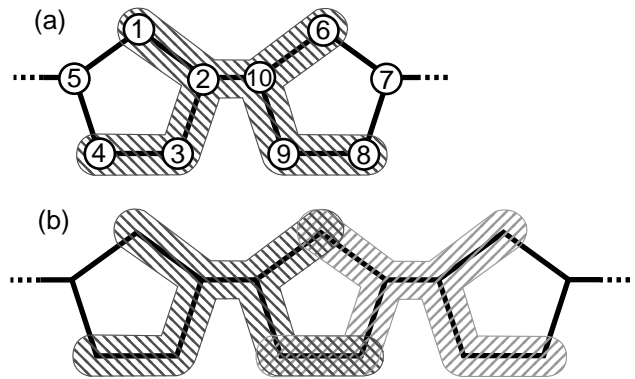}
\end{center}
\caption{Schematic view of (a) the eigenfunction that satisfies the
 connectivity condition and (b) the overlap of the adjacent
 eigenfunctions.}  \label{Eigen}
\end{figure}

Let us first examine the simple tight-binding model,
\[
H = -t \sum_{\langle i,j \rangle} c_i^\dag c_j + \varepsilon_0
\sum_{i=2}^5 n_i +\varepsilon_1 n_1,
\]
for a chain of five-membered rings, where we first assume that all the
transfer integrals within the ring and those connecting rings have the
same $t$, all the on-site energies are the same ($\varepsilon_0=0$),
except for the one at the top of the ring ($\varepsilon_1$).  We first
consider the half-filled case where one electron per site on
average. This is a reasonable assumption when we consider $\pi$-orbital
networks on such polymers.

Let us start from an observation that Mielke-Tasaki's condition is
satisfied when $\varepsilon_1=t$ ($>0$), a not unrealistic condition.
In this case, the third band with $E=0$, i.e., the half-filled band, has
no dispersion.  This provides a heuristic example for the Mielke-Tasaki
eigenfunctions: one can construct the most compact eigenfunction which
consists of two rings as depicted in Fig.~\ref{Eigen}(a), where the
amplitudes of the numbered sites are given as (1, 1, 0, -1, 0, -1, 0, 1,
0, -1).  The fact that the amplitudes at the sites 5 and 7 are zero
ensures the localized nature of this eigenfunction. Note that we are not
displaying a part of a Bloch function, but the whole eigenfunction. One
can also construct another, linearly-independent eigenfunction by simply
shifting the wave function from one ring to the next. The eigenvalue of
the new function is the same ($E=0$). So the set of all these
eigenfunctions can be a basis for the flat band.  Because the
eigenfunction is not confined in a single ring (unit cell) but extends
over the two rings, two neighboring bases (eigenfunctions) always
overlap with each other as shown in Fig.~\ref{Eigen}(b).  One cannot
remove such overlaps between the bases no matter how the linear
combination of these bases are taken.  This overlap is the origin of the
ferromagnetic coupling between the electrons in the flat band.

\subsection{Known polymers of five-membered rings}

\begin{figure}
\begin{center}
\includegraphics{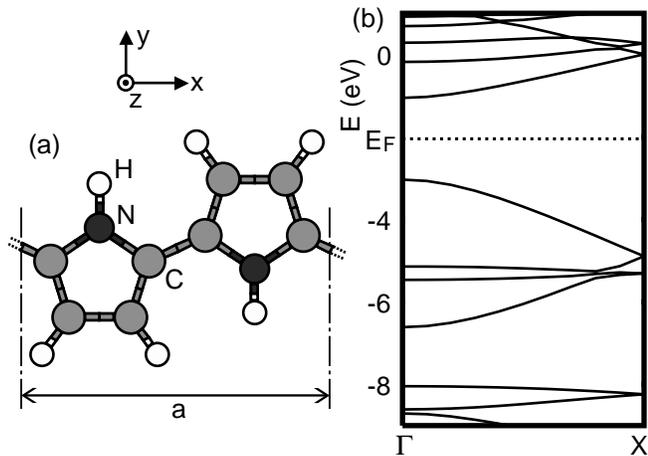}
\end{center}
\caption{(a) An atomic structure of polypyrrole. The structure extends
 along the right and left directions periodically. (b) The (GGA-DFT)
 band structure of polypyrrole.}  \label{NH}
\end{figure}

A typical polymer of five-membered rings is polypyrrole.  The atomic
structure and calculated band structures are shown in Figs.~\ref{NH}(a)
and (b), respectively. The five-membered rings alternate their
directions in this compound, so that the unit cell contains two rings.
X point in Fig.~\ref{NH}(b) has $\bm{k}=(\pi/a,0,0)$, where $a$ is the
lattice constant of this unit cell as shown in Fig.~\ref{NH}(a).
Because of the doubled unit cell, most of the bands are folded at X. For
directions ($y,z$) perpendicular to the chain, we have taken a cell size
large enough to avoid inter-chain interactions.  We can see that
polypyrrole has, not surprisingly, no flat bands around the Fermi
energy.

We have also looked at other typical polymers of five-membered rings,
including polythiophen (where an N-H block in polypyrrole is replaced
with an S atom) and polytriazole (where all the C-H blocks in
polypyrrole are replaced with N atoms). However, their band structures
have no flat bands, either.

\subsection{Designing the polymer}

This is exactly where the {\it designing} comes in.  Our strategy
consists of two approaches.  One is to replace H atom bound to N atom at
the top of the ring with various kinds of bases. According to the
tight-binding calculation described in the previous subsection, the
on-site energy ($\varepsilon_1$) of the top of the ring should be higher
than that of C atoms to have a flat band.  Because the on-site energy of
N is expected to be lower than that of C, the first attempt should be
the replacement of the N atom with other elements with higher on-site
energies. However, that should be difficult if one wants to retain the
existence of the $\pi$-orbital and single electron occupation on it.
Therefore, we have opted for controlling the on-site energy by replacing
the H atom with other functional groups, by leaving the N atom intact.

Alternatively, we can replace C-H blocks on the bottom edge of the
pentagons (3, 4, 8, 9 in Fig.~\ref{Eigen}(a)) with N atoms, where the
lowering of the on-site energies of the bottom of the ring may
effectively realize the required condition.  In other words, we use
polypyrrole and polytriazole as the platform to modify either the top or
the bottom of the ring.  Since it is not obvious which approach should
be better, we consider both platforms for all the substituents
considered here.

Substituents we have tested are sodium (Na), potassium (K), chlorine
(Cl), fluorine (F), cyanogen (CN), nitro (NO$_2$), sulfate (SO$_4$),
carboxyl (COOH), amino (NH$_2$), methyl (CH$_3$), and hydroxyl (OH).
For all these candidates, we have performed first-principles (GGA-DFT)
optimization of atomic structures, and obtained their band
structures. Most of them have no flat bands around $E_F$, but some of
them have turned out to have flat bands.  For these we have performed
further (GGA-SDFT) first-principles optimizations under the doping
condition to make the flat band half-filled.

\begin{figure}
\begin{center}
\includegraphics{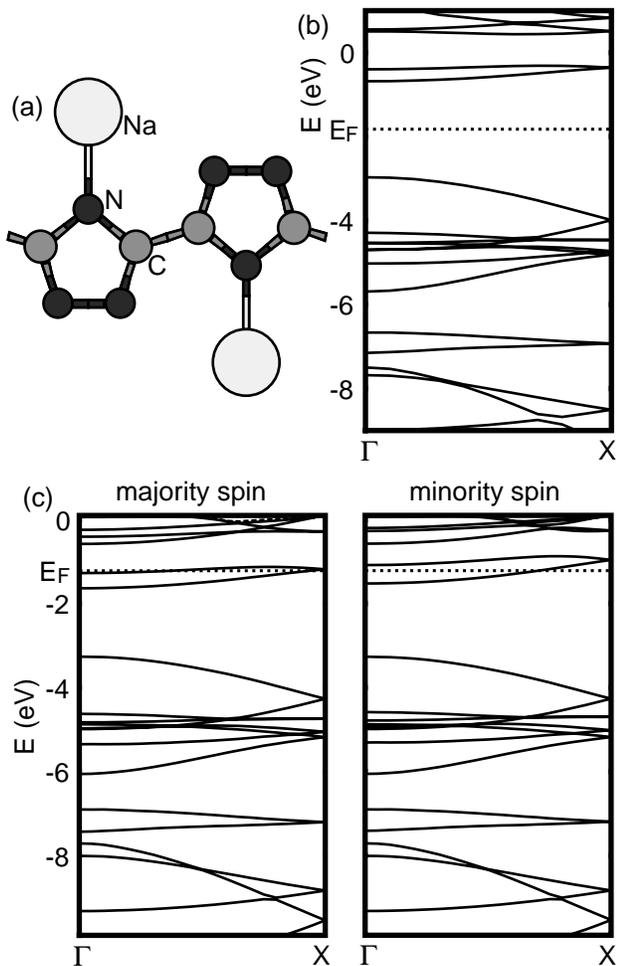}
\end{center}
\caption{(a) The optimized atomic structure of poly-sodium-triazole, the
 band structure of poly-sodium-triazole when undoped (spin
 unpolarized)(b), and doped (spin polarized)(c).} \label{NNA}
\end{figure}

First, we have examined single atoms (alkali metal or halogen) as
substituents.  This is motivated by the fact that small substituents may
be better than larger ones, because a bulky group at the top of the ring
may introduce a coupling (transfer) with the bottom of the adjacent
ring, which would destroy the basic assumption that the adjacent rings
are connected by one bond to satisfy Mielke-Tasaki's condition.

In order to raise the on-site energy of the top of the ring, we have
first considered an alkali-metal atom, Na, because of its low electron
affinity.  Figure~\ref{NNA}(a) shows the atomic structure of
poly-sodium-triazole, where hydrogen atoms in polytriazole are replaced
by sodium atoms. Figure~\ref{NNA}(b) shows the band structure, where one
can see that a nearly flat band exists above the Fermi energy.  Absence
of any unnecessary bands between the flat band and $E_F$ is good sign,
because we can dope carriers into the flat band without any
complications.

We have also examined the Na-substituted polypyrrole, and obtained a
similar band structure. However, in this case the flat band above $E_F$
is much closer to a dispersive band below $E_F$, which can introduce
unwanted inter-band interactions.  As for the alkali-metal we have also
considered potassium, but the case of Na has turned out to be better.

We move on to the doping.  First, the flat bands always appear as a pair
of bands folded at X point, except for the case of antiferromagnetic
ordering.  Therefore, we make the flat-band half-filled by doping two
carriers per unit cell into the pair of flat-bands.  Namely, one carrier
is doped for each ring. In poly-sodium-triazole, we have considered the
doping of two electrons per unit-cell. Here we realize the doping
condition by increasing the number of electrons with a uniform positive
background charge for charge neutrality.

After a geometrical optimization of doped poly-sodium-triazole, we have
calculated the spin-dependent (GGA-SDFT) band structure in
Fig.~\ref{NNA}(c).  One can see that the number of occupied states
differs between major- and minor-spins, which means that the system is
ferromagnetic.  It should be noted that (the pair of) the flat band is
not entirely spin-split. The difference of the number of spins is 0.65
per unit cell. The spin polarization is smaller than the expected
polarization (2.0).  We also found that the total energy (per unit cell
throughout this paper) of the ferromagnetic (F) state is lower by 12 meV
than that of the paramagnetic (P) state. We could not find
antiferromagnetic (AF) solution even when we started geometry
optimization from an AF electronic state.  So the F state is the most
stable in this material, but is only slightly lower in energy than the P
state.

The reason why this material gives such an insufficient result in spite
of the existence of the nearly flat band should be sought in the nature
of the flat-band's wave function.  We found that the half-filled flat
band here is made of an orbital localized around the Na atom, so does
not satisfy the local connectivity condition.  This should be why the
stability of the F state is very weak, while the finite
spin-polarization in GGA-SDFT should be a narrow-band effect rather than
a much more robust Mielke-Tasaki effect.  Indeed, the difference between
the weak ferromagnetism here and a robust ferromagnetism in
poly-aminotriazole, shown later, is a good demonstration of the
importance of the local connectivity condition, hallmark of the flat
band ferromagnetism.

\begin{figure}
\begin{center}
\includegraphics{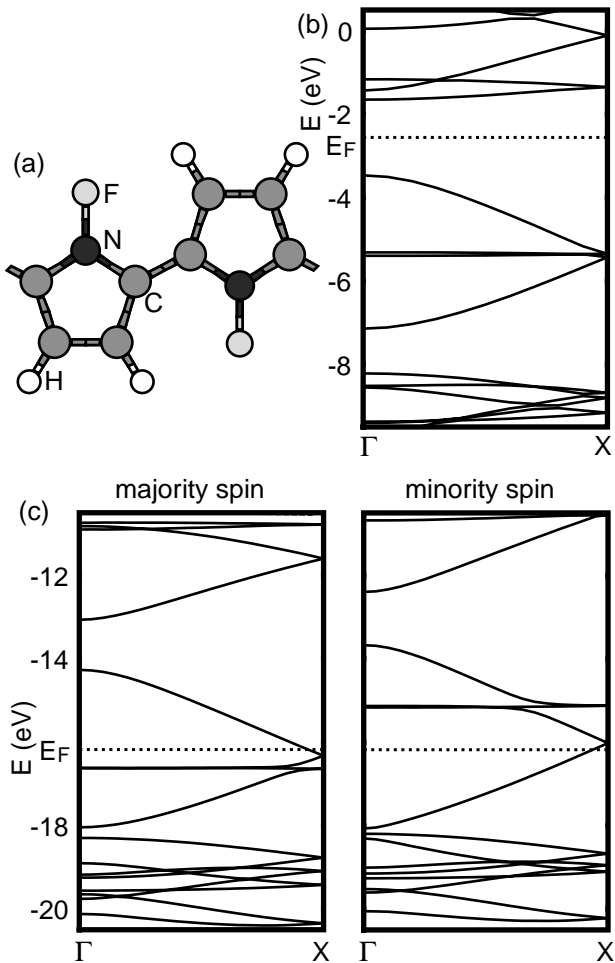}
\end{center}
\caption{(a) The optimized atomic structure of poly-fluoropyrrole, the
 band structure of poly-fluoropyrrole when undoped (spin
 unpolarized)(b), and doped (spin polarized)(c).} \label{NF}
\end{figure}

Now we digress a bit, and consider the contrary case of an atom having
high electron affinity such as a halogen, e.g., fluorine atom, which
might be heuristic.  Figure~\ref{NF}(a) shows the atomic structure of
poly-fluoropyrrole, where the fluorine atoms are substituted for
hydrogen atoms bound to nitrogen atoms in
polypyrrole. Figure~\ref{NF}(b) is the band structure obtained by
spin-unpolarized (GGA-DFT) calculation. We find a flat band which lies
just in between two dispersive bands below $E_F$.  Because of this the
doping greater than one carrier per unit cell is required.

So we consider making the flat-band half-filled by
doping four holes per unit-cell. Here, two holes are necessary to make
the dispersive band just below the Fermi level empty, and two holes are
necessary to make the folded flat bands half-filled.  Figure~\ref{NF}(c)
shows spin-dependent band structure after the geometry optimization
under the doping condition.  The system is seen to be ferromagnetic.
The difference of the number of spins per unit cell is just 2.0.  One
problem of this material, however, is that doping twice as large is
necessary.  When the doping is performed in this material, the flat band
shifts upward (downward) for minority (majority) spin.  Replacing the
fluorine atom with a chlorine atom results in a band structure similar
to the case of fluorine.  In fact an AF state, obtained by starting
geometry optimization from an antiferromagnetic state, has a total
energy 21 meV lower than that of F state.

We move on to functional groups as substituents to search further
possibilities.  We have considered an example, poly-hydroxytriazole,
where H atom in polytriazole is replaced by OH with a low electron
affinity. Although a pair of slightly dispersive bands exist just below
$E_F$, these bands, when doped with two holes per unit cell, become
mixed with dispersive bands underneath. As a result, the difference of
the number of spins in the F state is only 0.39 per unit cell, and the
paramagnetic state is more stable than the F state by 3 meV.

\begin{figure}
\begin{center}
\includegraphics{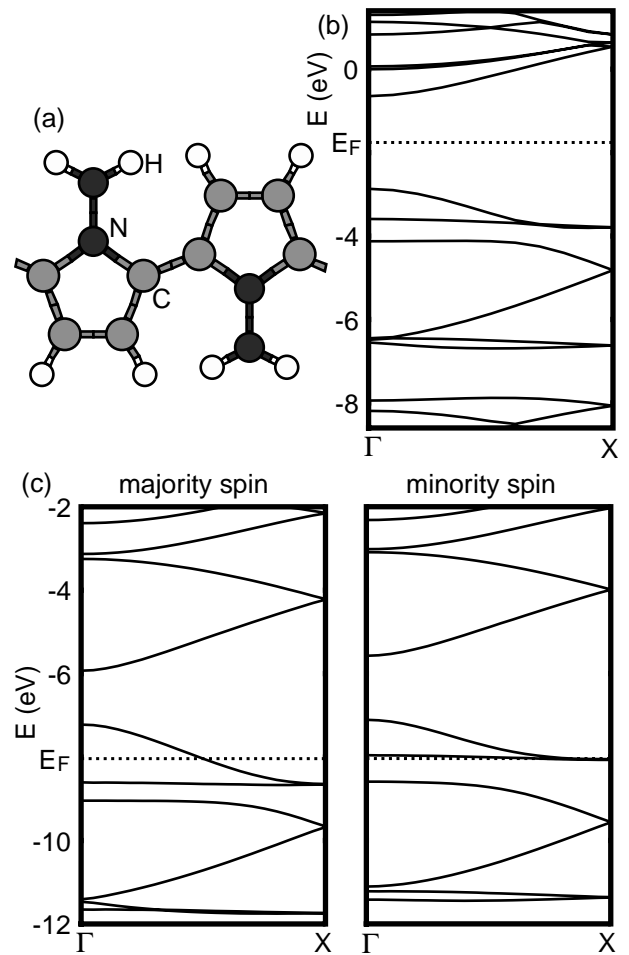}
\end{center}
\caption{(a) The optimized atomic structure of poly-aminopyrrole, the
 band structure of poly-aminopyrrole when undoped (spin unpolarized)(b),
 or doped (spin polarized)(c).} \label{CCNNH2}
\end{figure}

Figure~\ref{CCNNH2}(a) is the atomic structure of poly-aminopyrrole,
where H atom bound to N atom in polypyrrole is replaced by NH$_2$, whose
electron affinity is low.  In the band structure, Fig.~\ref{CCNNH2}(b),
one can find a pair of flat bands below $E_F$. The lower half of it is
very flat while upper half is slightly dispersive.

We have then doped two holes per unit cell. The band structure after
geometry optimization is shown in Fig.~\ref{CCNNH2}(c).  The difference
of the number of spins is 1.0 per unit cell.  As one can see from
Fig.~\ref{CCNNH2}(c), the Fermi level intersects the middle of the pair
of flat bands.  Although the lower part of the pair of flat bands is
quite flat, the dispersion of the upper half part is too large.  Here
only lower half part seems to work as a flat-band.  The total energy of
paramagnetic state in this material is found to be 65 meV higher than
that of ferromagnetic state, while an AF solution does not exist, so the
F state is the most stable.

All other substituents tested here except for the one in the next
section have turned out to be inappropriate.  In most of those
materials, the band which we expect to be flat is either too dispersive
or too deep.  For the remaining ones, band structures are qualitatively
similar to the examples described in this section. Poly-aminotriazole,
discussed in the next section, also gives qualitatively similar results
to poly-aminopyrrole shown in this section, but quantitatively much
better.

\section{Poly-aminotriazole}
\label{SecPAT}

\subsection{First principles calculation}

\begin{figure}
\begin{center}
\includegraphics{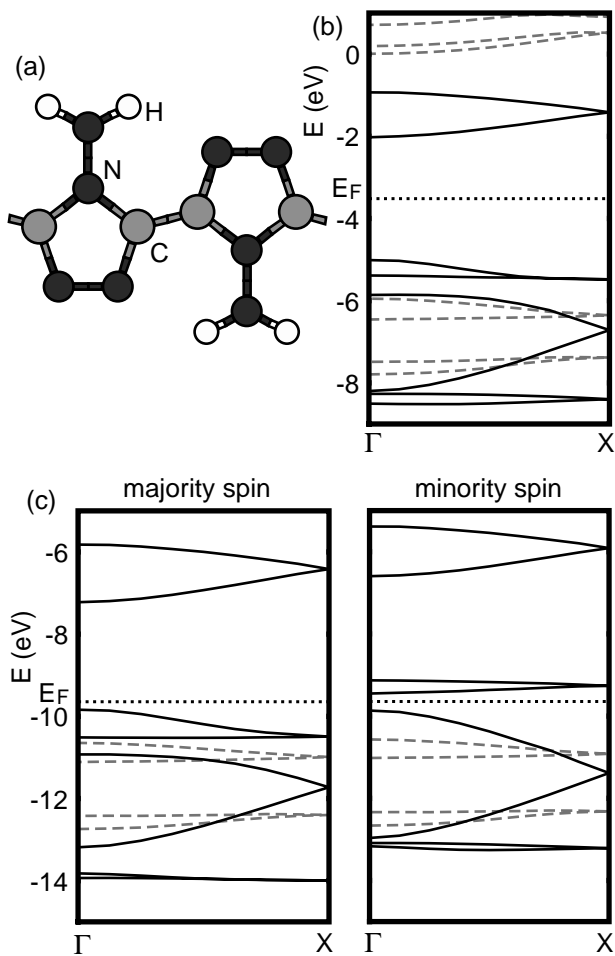}
\end{center}
\caption{(a) The optimized atomic structure of poly-aminotriazole, the
 band structure when undoped (spin unpolarized)(b), and doped (spin
 polarized)(c).  The bands having $\pi$ ($\sigma$) character are shown
 by solid (dashed) lines, respectively.  } \label{NNH2}
\end{figure}

The best candidate for ferromagnetic organic polymer among the materials
we have considered is poly-aminotriazole\cite{Arita1}.
Figure~\ref{NNH2}(a) shows the atomic structure of poly-aminotriazole,
where H atom in polytriazole is replaced with NH$_2$. The difference
between poly-aminopyrrole shown in the last section and this material is
whether the platform is polypyrrole or polytriazole, i.e., the C-H
blocks in the bottom of the ring is replaced with N atoms.  The band
structure of the poly-aminotriazole, shown in Fig.~\ref{NNH2}(b), is
similar to that of Fig.~\ref{CCNNH2}(b), but the dispersion of the upper
half of the pair of flat bands is smaller and the separation from
dispersive bands below is greater than that of
Fig.~\ref{CCNNH2}(b). These features are desirable for the flat-band
ferromagnetism.

After the optimization of the atomic structure under the doping
condition of two holes per unit cell, the spin dependent band structure
was calculated and shown in Fig.~\ref{NNH2}(c).  The result hits on the
ideal situation --- The pair of flat bands is made half-filled, and as a
result, fully occupied by majority spins while totally unoccupied by
minority spins. Reflecting this situation the difference of the number
of spins is the desired 2.0 per unit cell.

\begin{figure}
\begin{center}
\includegraphics{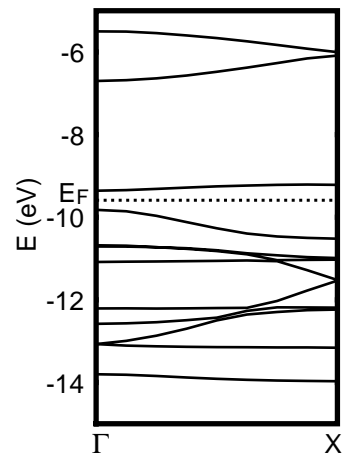}
\end{center}
\caption{A band structure for a metastable antiferromagnetic state in
poly-aminotriazole.} \label{AFband}
\end{figure}

In this material, we found a metastable antiferromagnetic state. The
band structure of the AF state is shown in Fig.~\ref{AFband}. This is
calculated with GGA-SDFT, but the resulting band structures for up- and
down-spins are the same, while the wave functions for up- and down-spins
are arranged alternately on the chain of rings.  This means that even
though we have avoided intra-ring AF instability by choosing five
membered rings, inter-ring AF can exist at least as a metastable state.
The upper and lower parts of the flat band folded at X point is
separated with $E_F$ in between.  The AF state is found to be 52 meV
higher in energy than the F state.  Paramagnetic state is even higher
(by 384 meV) than the F state. Therefore F state is the most stable in
this material.

\begin{figure*}
\begin{center}
\includegraphics{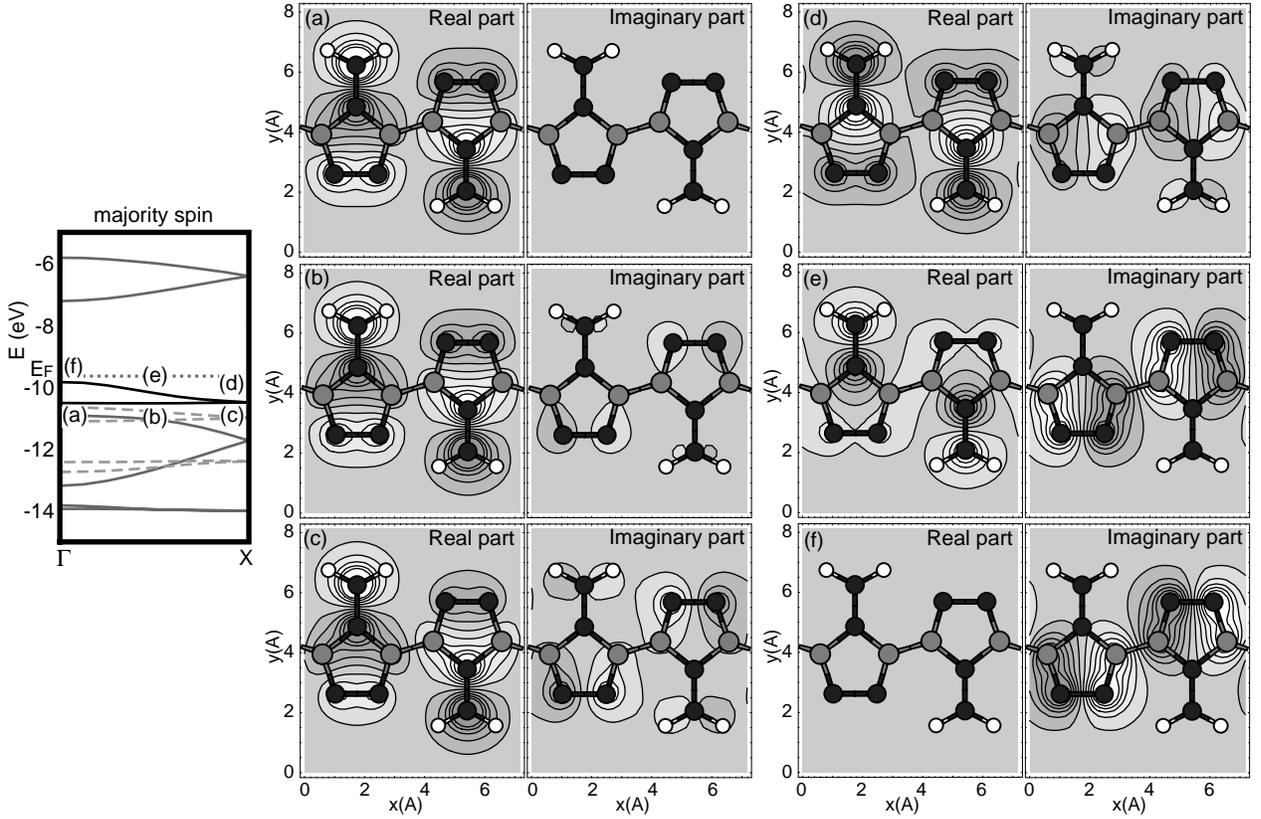}
\end{center}
\caption{Wave functions of the polarized lower band (a, b, c) and upper
 band (d, e, f) of poly-aminotriazole. (a, f) for $k_x=0$ ($\Gamma$),
 (b, e) for $k_x=\pi/2a$, and (c, d) for $k_x=\pi/a$ (X). Inset at the
 left side indicates the band and $k$-point of each figure.}
 \label{WavfFP}
\end{figure*}

Let us discuss how the flat-band ferromagnetism is achieved in this
material by combining the wave functions of the flat band obtained by
first principles calculation with the calculation in the tight-binding
model.  Figure~\ref{WavfFP} shows the majority-spin's wave functions in
the pair of flat bands just below the Fermi level in Fig.~\ref{NNH2}(c).
While the phases of the wave functions can be taken arbitrarily, we
choose for clarity the wave functions of the lower and upper bands at
$\Gamma$ point (Figs.~\ref{WavfFP}(a) and (f)) to be real and pure
imaginary, respectively, and those at intermediate $k$-points to be
continuous.  If we first look at the real part, typically for
Fig.~\ref{WavfFP}(a), the wave function for each monomer (aminotriazole)
has two nodal lines which divide the amplitude into three parts,
corresponding to NH$_2$, C-N-C, and N-N blocks.  For the imaginary part,
typically for Fig.~\ref{WavfFP}(f), the wave function has one nodal
line, which divides the amplitude into two, consisting of two C-N
blocks.  The two wave functions (real and imaginary), which are
orthogonal, are mixed at the intermediate $k$-points.

\subsection{Tight-binding model}

\begin{figure}
\begin{center}
\includegraphics{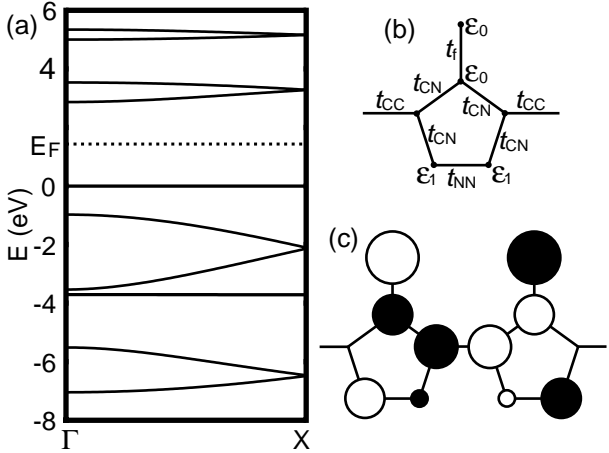}
\end{center}
\caption{(a) Band structure in the tight-binding model. (b) Tight-binding
 parameters used for the fitting. (c) The eigenfunction which satisfies
 the connectivity condition.}  \label{TBfit}
\end{figure}

In order to capture the essence of the flat band, here we introduce a
tight-binding model which represents $\pi$-orbital's network in this
system.  Figure~\ref{TBfit}(a) shows its band structure.  The parameters
of the model displayed in Fig.\ref{TBfit}(b) are fitted so as to
reproduce Fig.~\ref{NNH2}(b). Obtained values are: $t_{\rm CN}= t_{\rm
CC} = t_{\rm f} = 2.5$ eV, $t_{\rm NN}= 3$ eV, $\varepsilon_0 =-1.4$ eV,
and $\varepsilon_1 = -0.5$ eV.  One can see that the tight-binding band
(Fig.~\ref{TBfit}(a)) excellently agrees with the $\pi$-bands obtained
by first-principles calculation (solid lines in Fig.~\ref{NNH2}(b)).

Figure~\ref{TBfit}(c) shows the tight-binding wave function of the
eigenstate corresponding to that in Fig.~\ref{Eigen}(a). In this
material, one site is added to the top of the ring which represents a
nitrogen atom of NH$_2$. The open and closed circles indicate the sign
of the wave function, respectively, and their sizes amplitudes. This
wave function satisfies the local connectivity condition.

Here it should be noted that the substitution of NH$_2$ for H not only
raises the on-site energy of the top of the ring but also introduces an
extra $\pi$-orbital of the N in NH$_2$. The existence of the extra site
completely changes the connectivity condition, so that we have no longer
to satisfy $\varepsilon_1=t > 0$, which has turned out in the present
study to be rather difficult condition to realize.

\begin{figure}
\begin{center}
\includegraphics{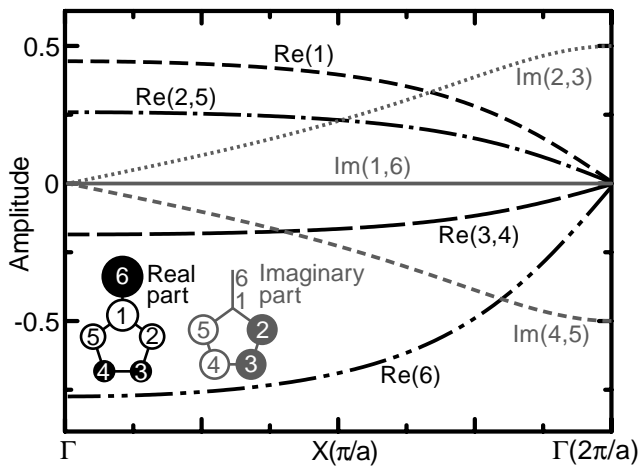}
\end{center}
\caption{$k$-dependence of the wave functions in the tight-binding model.}
 \label{WavfTB}
\end{figure}

Figure~\ref{WavfTB} shows $k$-dependence of the tight-binding wave
functions of the flat band. Here $k$-space is extended to the second
Brillouin zone, so the left and right halves of the figure represent the
wave functions of the lower and upper parts of the pair of flat bands,
respectively.  Phase factors are chosen again for clarity as in
Fig.~\ref{WavfFP}. The label ``Re(2,5)'' indicates that the line
represents the real part of the wave function at the site 2 and 5, where
the numbering of the sites is shown in the insets. On the
$\Gamma$-points at the left- and right-edge of the figure, the wave
function becomes real and pure imaginary, respectively.  These two wave
functions are displayed in the inset of Fig.~\ref{WavfTB}, where it
should be noted that we display a part of the periodically extending
Bloch functions, in contrast to Fig.~\ref{TBfit}(c) which shows the
entire eigenfunction.

\begin{figure}
\begin{center}
\includegraphics[width=6cm]{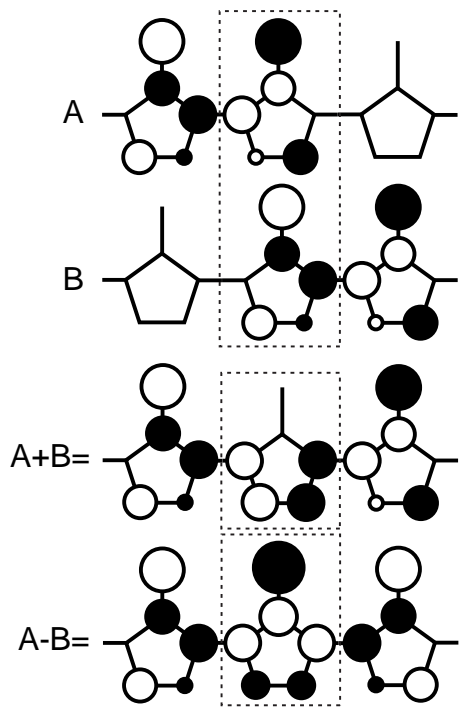}
\end{center}
\caption{Schematical linear combinations of the neighboring
eigenfunctions.}  \label{TBmath}
\end{figure}

One can see that the symmetry and the character of the wave functions
shown in Fig.~\ref{WavfTB} is similar to the first-principles wave
function (Fig.~\ref{WavfFP}).  These wave functions can be constructed
by a linear combination of the eigenfunctions shown in
Fig.~\ref{TBfit}(c).  Figure~\ref{TBmath} illustrates how the linear
combination of the localized eigenfunctions provides the Bloch function.
When the two neighboring eigenfunctions, A and B, are added, the sum at
the ring on which they overlap (enclosed by dotted lines) forms the
shape of the imaginary-part wave function as shown in the inset of
Fig.\ref{WavfTB}. When subtracted, it forms the shape of the real-part
wave function. The entire imaginary- and real-part wave
functions are constructed by the summation of all localized
eigenfunctions (A+B+C+D+$\cdots$) and the combination of the summation
and subtraction (A$-$B+C$-$D+$\cdots$), respectively.  The real-part wave
function and the imaginary-part wave function are orthogonal to each
other.  The gradual mixing of these two orthogonal wave functions as $k$
is changed is the very character of the Mielke-Tasaki flat band, which
should never occur on an ordinary flat band.

\subsection{Comparison with the Hubbard model}

\begin{figure}
\begin{center}
\includegraphics[width=7.5cm]{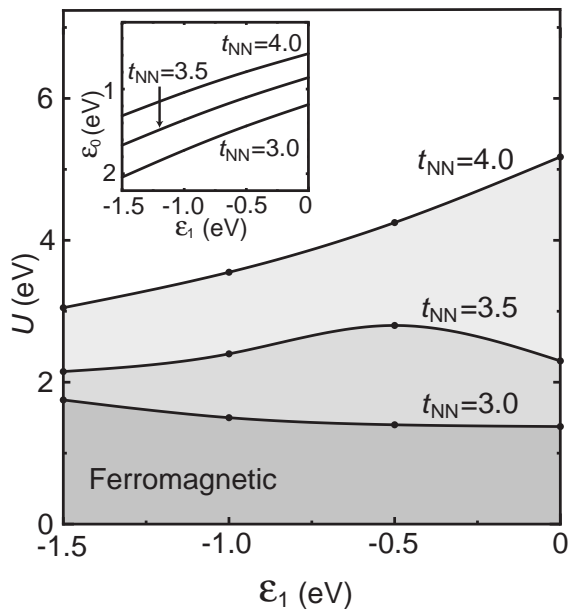}
\end{center}
\caption{Stability of the ferromagnetic phase calculated with an exact
 diagonalization of the Hubbard model.}
 \label{Hubbard}
\end{figure}

Having constructed the tight-binding modeling we proceed to the question
of whether the ground state is spin-polarized in the presence of the
Hubbard interaction, ${\cal H}_{U}=U \sum_{i}
n_{i\uparrow}n_{i\downarrow}$.  In Fig.~\ref{Hubbard} we show the phase
diagram against $U$ and $\varepsilon_1$ obtained with an exact
diagonalization calculation for a 12-site (2 rings) Hubbard model for
$t_0=2.5$ eV and various values of $t_{\rm NN} = 3.0 - 4.0$ eV.  As
indicated in the inset, $\varepsilon_0$ is chosen throughout to satisfy
the condition for the flat band,
\begin{eqnarray*}
\varepsilon-\varepsilon_1&=&(1-\varepsilon)
(t_{\rm NN}^2-(\varepsilon-\varepsilon_1)^2)-t_{\rm NN}, \\
(\varepsilon-\varepsilon_1+t_{\rm NN})/(1-\varepsilon)
&=&-t_{\rm NN}/(\varepsilon_0-\varepsilon)
-t_{\rm NN}(\varepsilon-\varepsilon_0),
\end{eqnarray*}
where $\varepsilon$ is the eigenenergy of the flat band and $t_{{\rm
CN}}= t_{{\rm CC}} =t_0$ ($=1$ here) is assumed for simplicity. We can
see that we have indeed a ferromagnetic phase unless the repulsion is
too strong (i.e., $U < U_c$ with $U_c = 2 \sim 5$ eV). Our preliminary
quantum Monte Carlo calculation confirms that two unit cell is
sufficient to roughly determine the ferromagnetic phase boundary.  
Peculiar destruction of the ferromagnetism above $U > U_c$ is 
another hallmark of the Mielke-Tasaki magnetism.\cite{Arita98}

\section{Deviation from the half-filled flat band}
\label{SecDiscuss}

\subsection{Comparison of the total energies}

In order to realize ferromagnetism in the materials we proposed here,
carriers have to be doped into the flat band.  We can consider several
possible methods to achieve this such as field effect doping\cite{Lee},
photo-induced carrier injection\cite{Hiroi}, as well as chemical
doping. The first two are now attracting much attention in molecular
electronics, but they are very recent techniques and may need further
developments.  We can instead concentrate on the chemical doping, where
the polymers are crystallized along with the accepters, and we found it
feasible. Details are presented elsewhere\cite{Arita2}.  Now, if the
flat-band ferromagnetism can occur not only just at the half-filling but
also away from that, this would make the chemical doping much easier.

We have to know how robustly the ferromagnetism survives when the
filling of the flat band deviates from just half-filled.  Although there
is a proof\cite{Mielke1} and a numerical study\cite{Sakamoto} for a
Mielke-Tasaki lattice which show that slight deviations do not wash out
the ferromagnetism, those conclusions may not be simply applied to other
models.  For the present model, calculations taking full account of the
electronic correlation such as the exact diagonalization of the Hubbard
model would be unsuitable, because changing the number of electrons in
small systems corresponds to too big a change in the filling.  We have
instead performed an SDFT calculation, because it can handle the
fractional number of electrons (per unit cell) if a uniform background
charge is introduced.

At this point we can note the following.  In SDFT calculations,
electronic correlations are not fully taken into account, so we can
question whether the flat band ferromagnetism may be described within
the SDFT framework.  This problem has been addressed in our previous
publication\cite{Arita1} where SDFT and exact diagonalization of a
Hubbard model have given qualitatively consistent results.  This is not
surprising, since one of the remarkable features of the flat-band
ferromagnetism is that the magnetism exists over the whole range
(infinitesimal $< U < \infty$) of the interaction as proved rigorously,
so the weak-$U$ case (for which the methods such as SDFT is meant to be
applicable) crosses continuously over to strong-$U$ case.  As far as the
on-site repulsion is not too strong, the comparison between the gain of
the exchange-correlation energy by spin polarization (F state) and the
gain of the band energy by forming AF order (or equivalently, a spin
density wave (SDW) with the periodicity of two five-membered rings)
accompanied by a lattice relaxation should predominantly determine which
state is stable.  These two energies can be accurately estimated by SDFT
calculations. Therefore, we expect here that SDFT can serve our purpose,
to see the band-filling dependence of the total energies of F, AF, and P
states.

\begin{figure}
\begin{center}
\includegraphics{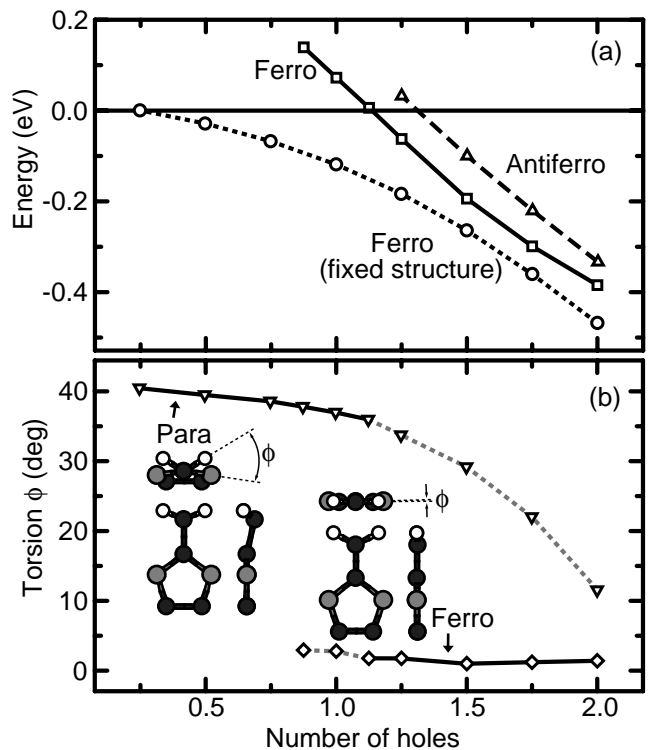}
\end{center}
\caption{Doping dependence of (a) the total energy (relative to the
paramagnetic state) of F(solid line), AF(dashed line), and F(fixed
structure, dotted line, see text) states measured from that of P state,
and (b) torsion angle of C-N-N-H in F(diamonds) and P(reversed
triangles) states.  Stable states are connected by a solid line, while
metastable states by a dotted line.  Insets show front, side, and top
views of the atomic structure of the aminotriazole in P and F states.}
\label{Doping}
\end{figure}

Figure~\ref{Doping} (a) shows band-filling dependence of the total
energies of the F and AF states of the poly-aminotriazole measured from
that of the P state. The position where the number of holes, $n_h$, is
2.0 corresponds to the half-filling.  The results for F, AF, and P are
obtained by fully relaxing the atomic geometry for each electronic
state. It can be seen that F state is stable in the range $1.1 < n_h
\leq 2.0$, while P state is stable for $n_h<1.1$. In the region where F
is stable, AF is found to be metastable.  In the P-region, $n_h<1.1$, no
AF state is found.  This result shows that the ferromagnetism in this
material is robust, even down to the quarter filling, against the
deviation from the half-filling. The result that quarter-filling is
acceptable greatly relaxes the difficulty in doping.

\subsection{Correlation with the atomic configuration}

Since the polymer proposed here has functional groups, we should also
examine how the three-dimensional atomic configuration is correlated
with the band-filling dependence.  Namely, all the atoms do not lie on a
plane: in poly-aminotriazole two hydrogen atoms in NH$_2$ in the undoped
polymer lie above the plane of the rings, while all the atoms are
co-planar in the half-filled poly-aminotriazole.  We can more closely
look at the torsion angle, $\phi$, defined as the angle with which
C-N-N-H atoms are connected.  Figure~\ref{Doping}(b) shows the torsion
angles in the P and F states, respectively.  From Figs.~\ref{Doping}(a)
and (b), one can see that the breakdown of the ferromagnetism with the
decrease in the number of holes is correlated with the change in the
atomic configuration.

When the system is half-filled ($n_h=2.0$), all atoms become co-planar,
where the N atom of the NH$_2$ base has $sp^2$ hybridized orbitals,
forming chemical bonds with two H atoms and a N atom at the top of the
ring.  The network of the $\pi$ orbitals in poly-aminotriazole is formed
by those of the rings and that of NH$_2$'s.  When the number of holes is
decreased and becomes $n_h<1.1$, the N atom of the NH$_2$ forms $sp^3$
hybridized orbitals, i.e., the $\pi$ orbital at NH$_2$ disappears.
Since the chemical bonds to the N atom and two H atoms are formed in the
three directions of $sp^3$, leaving one lone pair, NH$_2$ has a
three-dimensional structure.  This time, the network consists of five
$\pi$-orbitals of the rings and a $sp^3$ hybridized orbital of the lone
pair. This change is disadvantageous for the scheme of the flat-band
formation (Fig.~\ref{TBfit}) due to the balance of the six $\pi$
orbitals.  This is considered why F state always takes a flat geometry
while P state a non-flat geometry.

To see what part of the energy affects the change in the stable state
when $n_h$ is decreased, we have calculated the energy difference
between the F and P states by fixing the atomic structure to the optimum
one for the F state at $n_h=2.0$.  The result is plotted as a dotted
line in Fig.~\ref{Doping}(a).  This shows that the value is negative
over the whole range plotted here, which indicates that the {\it
electronic} part of the total energy always favors F state regardless of
the filling of the flat band.  Therefore, the change in the stable state
at $n_h=1.1$ can be ascribed to the change in the structure, especially
of NH$_2$.  If we can find a material in which there is no such change
in the atomic structure but the parameter values are similar, we can
expect an occurrence of the ferromagnetism even for a low-doping region
less than quarter-filling.

In summary, the present work describes an example of materials design
from first principles for the purpose of realizing flat-band
ferromagnetism in organic polymers.  We have found a promising material,
polyaminotriazole, for which why it is the best and the robustness of
the ferromagnetism has been discussed.

\begin{acknowledgments}
We would like to thank T. Hashizume, B. Choi, M. Ichimura and
J. Yamauchi for fruitful discussions.  We are also indebted to
H. Nishihara and Y. Yamanoi for discussions on the chemistry of
polymers.  This study was performed through Special Coordination Funds
for Promoting Science and Technology of the Ministry of Education,
Culture, Sports, Science and Technology of the Japanese Government. The
first-principles calculations were performed with TAPP (the Tokyo
Ab-initio Program Package).
\end{acknowledgments}

\end{document}